\documentclass[pre,twocolumn,amsmath,amssymb,floatfix]{revtex4}

\usepackage{graphicx}
\usepackage{subfigure}
\usepackage{dcolumn}
\usepackage{epsfig}
\usepackage{amssymb}
\usepackage{amsmath}
\usepackage{rotating}
\usepackage{color}

\newcommand{\gammap}{\dot{\gamma}}

\begin{document}

\title{Microfluidic velocimetry reveals spatial cooperativity in soft glassy flows}

\author
{J. Goyon,$^{1}$ A. Colin,$^{1,\ast}$ G. Ovarlez,$^{2}$ A. Ajdari,$^{3}$ L. Bocquet, $^{4,5,\ast}$ \\
\normalsize{$^{1}$ LOF,  Universit\'e Bordeaux 1, UMR CNRS-Rhodia-Bordeaux 1 5258,}  \\
\normalsize{33608 Pessac cedex, France}\\
\normalsize{$^2$ LMSGC, Universit\'e Paris Est, Institut Navier, 77420 Champs sur Marne, France}\\
\normalsize{$^3$ UMR CNRS Gulliver 7083, ESPCI, 75005 Paris, France}\\
\normalsize{$^4$ Laboratoire PMCN, Universit\'e Lyon 1; Universit\'e de Lyon,}\\
\normalsize{UMR CNRS 5586, 69622 Villeurbanne, France} \\
\normalsize{$^5$ Physics Department, Technische Universit\"at M\"unchen, 85748 Garching, Germany}
\\
%\normalsize{$^\ast$To whom correspondence should be addressed;}\\
\normalsize{E-mail:  lyderic.bocquet@univ-lyon1.fr, annie.colin-exterieur@eu.rhodia.com}
}
\date{\today}

\maketitle

{\bf Amorphous glassy materials of diverse nature -- concentrated emulsions, granular materials, pastes, molecular glasses --
display complex flow properties, intermediate between solid and liquid, which are at the root of their use in many applications \cite{Jop,Debenedetti,Biroli}.
A classical feature, well documented yet not really understood, is the very non-linear nature 
of the flow rule relating stresses and strain rates \cite{Coussot,Fuchs}. 
Using a microfluidic velocimetry technique, we characterize the flow of thin layers 
of concentrated emulsions, confined in gaps of different thicknesses by surfaces of different roughness.
Beyond the classical non-linearities of the rheological behaviour, we evidence finite size effects in the flow behaviour
and the absence of an intrinsic local flow rule.
In contrast, a rather simple non-local flow rule is shown to account for all the velocity profiles. 
This non-locality of the dynamics is quantified by a length, characteristic of the cooperativity of the flow 
at these scales, that is unobservable in the liquid state (lower concentrations) and that increases 
with concentration in the jammed state. Beyond its practical importance for applications involving thin layers, e.g. coatings, 
our assessment of non-locality and cooperativity echoes observations on other glassy, jammed and granular systems, 
suggesting a possible fundamental universality.}
\vskip 1cm

Glasses and jammed matter are at the heart of many challenging questions in condensed matter physics. 
The glass transition and the physical and mechanical properties of materials in the glassy or jammed state, from granular materials to dense colloidal suspensions, remain puzzle to a large extent \cite{Debenedetti,Goldenberg,Ostojic,Biroli,Liu,Coussot, Fuchs}. 
%ADD
The  transition itself is characterized by a dramatic slowing down of the system, associated with a dynamical arrest of the microscopic structure \cite{Biroli}.
%ADD
While a large amount of work has been dedicated in recent years to demonstrating and characterizing dynamical heterogeneities in granular, glassy and glass forming materials \cite{Berthier2004a,Weeks,Berthier2005,Gao,Priestley,Dauchot2005,Pouliquen2004,Durian2007}, the question of whether and how these dynamical heterogeneities influence the flow behaviour remains \cite{Varnik,Leonforte,Isa}. Indeed one expects that such cooperative effects, which are in essence dynamical processes, should manifest themselves in the flow dynamics of jammed materials. This question is essential from the perspective of both fundamental science and applications: flows of jammed and glassy materials and glassy films are ubiquituous in nature and industry, {\it e.g.} in granular flows, coatings with thin polymer films, lubrifying process in solid friction, 
in the food and cosmetic industries, and even in pedestrian dynamics
\cite{Helbing}.

%In the context of granular materials, a milestone has been achieved recently with the development of constitutive laws able to describe granular flows in various geometries\cite{Jop}. However, non-local rheological effects observed in granular flows close to the jamming transition are not captured by this approach \cite{Andreotti}. 
In the present work, we aim to develop a constitutive law governing flows of jammed materials. 
We take advantage of a local velocity measurement technique to follow
the local flow behaviour of a film of a confined soft glassy material (here a concentrated emulsion).
We demonstrate the existence of {\it finite-size effects in the flow dynamics} that cannot be apprehended within our current understanding of the behaviour of yield stress fluids.
%These flow cooperativity effects lead to a breakdown of the local relationship between the shear rate and shear stress. We introduce a non-local constitutive law that allows all velocity profiles and flow characteristics to
%be reproduced quantitatively.
%A flow cooperativity length emerges naturally, whose behavior differs from the heterogeneity length of mobile regions measured in spontaneous fluctuations  \cite{Weeks,Berthier2005}. We argue that this novel characteristic length measures the zone of influence of the plastic events involved in the flow of these materials.

As schematized in Fig. 1, we probe the flow of a jammed emulsion in various geometries and confinements: shear or pressure driven planar flow; wide gap Couette cell (one centimer gap) or narrow microchannel (from tens to hundreds of microns in width).
%ADD
The emulsion is made of silicone droplets ($\approx 6.5\mu$m in diameter) in an index-matched glycerine-water mixture (Supplementary Information, Method).
%ADD
We are specifically interested in the {\it local flow curves} of the material, which relate the local shear stress $\sigma$ to the local shear rate $\dot\gamma$, Fig. 1C-D. Such curves can be
constructed from the velocity profiles measured 
 in the jammed emulsion in both geometries using local velocimetry techniques (Supplementary Information, Method), see insets of Fig.~1C-D.
%From these profiles, it is possible to construct
%%We first focus on 
%{\it local flow curves} of the material by relating the local shear stress $\sigma$ versus the local shear rate $\dot\gamma$.
%This property is accessible to these experiments by measuring local velocity profiles (emulsion velocity as a function of position), see Method.
Indeed in both geometries the stress distribution is known from mechanical equilibrium: $\sigma(R)= \frac {\Gamma}{2\pi H R^2}$ at a position $R$ in the Couette cell (of height $H$) under a torque $\Gamma$; $\sigma(z)=\frac{\Delta P}{L} (z-{w\over 2}) $, at a position $z-{w\over 2}$ from the centerline of the microchannel, under an applied pressure drop $\Delta P$ between the two ends of the channel of length $L$ \cite{Degre2006}. 
%Note that slippage effects at the wall surface do not alter this relationship.
Local shear rates $\dot\gamma$ are deduced directly from the velocity profiles. As shown in Figs.1.C and 1.D, the local flow curves exhibit very different behaviours depending on the
confinement of the material. In the {\it wide}Ê
gap Couette cell, all local flow curves obtained for different torques show a perfect superposition (Fig 1.C). The flow curve is furthermore well described by a
classical Herschel-Bulkley (HB) model,  $\sigma= \sigma_0+A \gammap^{1/2}$, with $\sigma_0$ a 'dynamical' yield stress and $A$ a material constant. 
On the other hand, in the {\it narrow}Ê
microchannels, data for different pressure drops $\Delta P$ are scattered in the whole figure and do not collapse on a single rheological curve (Fig 1.D), see also Supplementary Figure SI-4.
Accordingly, while a single HB model accounts perfectly for all the velocity profiles in the wide gap Couette cell (inset of Fig.1.C), {\it such is not the case for % {\it it was not possible to fit the various velocity profiles} in 
the narrow microchannels} (Supplementary Figure SI-8).
%on the basis of a unique HB rheological model 
%(not shown). 
Furthermore, {\it below the jamming point}, $\phi<\phi_c$, the local HB rheological model is fully able to reproduce the flow behaviour, even in narrow channels (Supplementary Figure SI-3).
In summary, a first key feature emerges from the experimental data: in the jammed state, $\phi>\phi_c$, {\it there is no universal local relationship between stress and shear rate}. 

As a further confirmation of this point, the overall shape of the velocity profiles, for the same wall shear stress,  is shown to vary with the thickness of the channel (Fig. \ref{courbesfittes}), demonstrating {\it finite-size effects in the flow properties} of the jammed emulsion. Rougher surfaces are shown to induce higher shear rates for the same confinement and shear stress at the wall.
We did furthermore verify that the above results do not stem from a structural change of the emulsion itself:
we checked that neither the size distribution of the droplets, nor their shape, was affected by the flow and in all the geometries no variation of local droplet concentration was measured within the gap (Supplementary Method).
%This quantity is directly measured by MRI in the Couette cell.  In microchannel flows, rhodamine was added to the continuous phase and the analysis of the fluorescence revealed no local variation of the measured intensity, suggesting no variation in the local droplet concentration. 
Altogether, our results suggest that the flow does not couple to a structural order parameter, 
so that non-local effects are 'intrinsic' to the rheology of the jammed emulsion, in line with  similar observations in numerical simulations for glassy flows \cite{Varnik}.

A second crucial feature is the {\it experimental proof of a specific surface rheology} at the channel walls, distinct from the bulk flow behaviour. 
First, as commonly observed in flows of concentrated systems, slip effects show up at the confining walls \cite{Cloitre}. We have characterized these effects by measuring the velocity slip-surface stress relationship (Fig. \ref{Slip}a), for 
%Confining surfaces with 
two different roughness (Supplementary Method and Supplementary Figure SI-2).
Strikingly, a remarkable collapse of the data is exhibited for various confinements and pressure drops, suggesting that this relationship does only depend on the surface roughness, see also Supplementary Figure SI-7a. 
However, a careful inspection of  Fig. \ref{courbesfittes} shows that beyond slip effects, the {\it shape}
itself of the flow profiles does depend on the surface characteristics of the confining wall,
%: for the same confinement and pressure drop, the maximum velocity of the emulsion (corrected for the slip velocity) is much higher for  the rough confining wall surfaces than for the smooth surfaces,
suggesting that the wall roughness makes the system more fluid. In an attempt to rationalize this behaviour we plot the wall shear rate as a function of the wall shear stress in Fig. \ref{Slip}b (see also Supplementary Figure SI-7b). Again a striking collapse of all data for various confinements and pressure drops is exhibited. We emphasize that such a behaviour is not expected {\it a priori} since the same wall stress can be achieved with different confinements and pressure drops.
%The remarkable behavior of both the slip velocity and surface flow curve strongly suggests the existence of a specific surface rheology characterizing the flow behavior at the surface.

On the basis of these experimental facts, 
we now propose a theoretical framework able to rationalize these finite-size effects in the rheological behaviour. 
In concentrated emulsions, and more generally in soft glasses, flow occurs via a succession of reversible elastic deformations and local irreversible plastic rearrangements  associated with a microscopic yield stress \cite{Princen,Schall,Picard2002,Picard2005}. These localized plastic events induce a non-local, long ranged, elastic relaxation of the stress over the system.  
%ADD
The number of plastic rearrangements per unit time, $\nu$, 
%which depends {\it a priori} on the shear rate, 
plays the role of an inverse relaxation time and therefore controls the flow and relaxation of the material: a higher rate is associated with a more fluid system. 
Up to an elastic modulus $G$, the rate of rearrangements $\nu$ is equivalent to a ``fluidity'', $f$, defined as $\sigma={1\over f} \gammap$, a quantity that has been used in recent studies to characterize the rheology of yield stress fluids \cite{Picard2002}.
%Writing the stress versus shear rate relationship 
%as $\sigma={1\over f} \gammap$, the quantity $f$, which will be denoted as a fluidity, can thus be rationalized as the ratio between the number of rearrangements per unit time $\nu$ divided by an elastic modulus 
%$G$ \cite{Picard2002}.
In the absence of any non-local effect, the fluidity would reduce to its {\it bulk} value, $f_{\rm bulk}=\gammap/\sigma_{\rm bulk}$
[note that for  the HB expression reported above,  $f_{\rm bulk}\propto \gammap$ in the quasistatic limit, as $\sigma\rightarrow \sigma_0$]. 
Now, due to the non-local elastic relaxation of plastic rearrangements ('dynamical cooperativity')\cite{Picard2005,Schall}, 
a dynamically active region will induce agitation of its neighbours, and thus a higher rate of plastic rearrangements (and vice-versa for a dynamically quiescent region).
This suggests that non-local effects will occur for the rearrangements rates and consequently for the fluidity, $f$.
In order to capture this physical picture, we assume that 
the local fluidity, $f(z)$, obeys a non-local equation in the steady state:
\begin{equation}
f(z)=f_{\rm bulk}+\xi^2 \frac{\partial^{2} f(z) }{\partial^{2}z} 
\label{diff}
\end{equation}
%ADD
In this equation, the non-local term describes how
the plastic activity spreads spatially over the system due to non-local elastic relaxation: it accounts for the cooperativity involved in the plastic events occuring during flow and
$\xi$ denotes a bulk 'flow cooperativity length'. 

Together with the definition $f(z)=\gammap(z)/\sigma(z)$, Eq. (\ref{diff}) constitutes our non-local rheological model for the emulsion. 
%
%We have solved Eq. (\ref{diff}) numerically, 
An analytical solution of Eq. (\ref{diff}) can be obtained, giving the fluidity at some position $z$ in terms of an integral of the shear-stress over the whole channel width. For convenience we solved Eq. (\ref{diff}) numerically
with $f (z)=\gammap_{\rm wall} / \sigma_{\rm wall} $ as a given boundary condition at the walls and 
$f=0$ in the centerline.
Velocity profiles are deduced by integrating the computed local shear rate $\gammap(z)$. Let us first note that {whatever the boundary condition at the wall} the solution of Eq. (\ref{diff}) can not reduce to its local form, $f_{bulk}(z)=f_{bulk}(\sigma(z))$, due to the diffusive, non-local term.

\begin{figure*}
\begin{center}
\scalebox{1.0}{\includegraphics {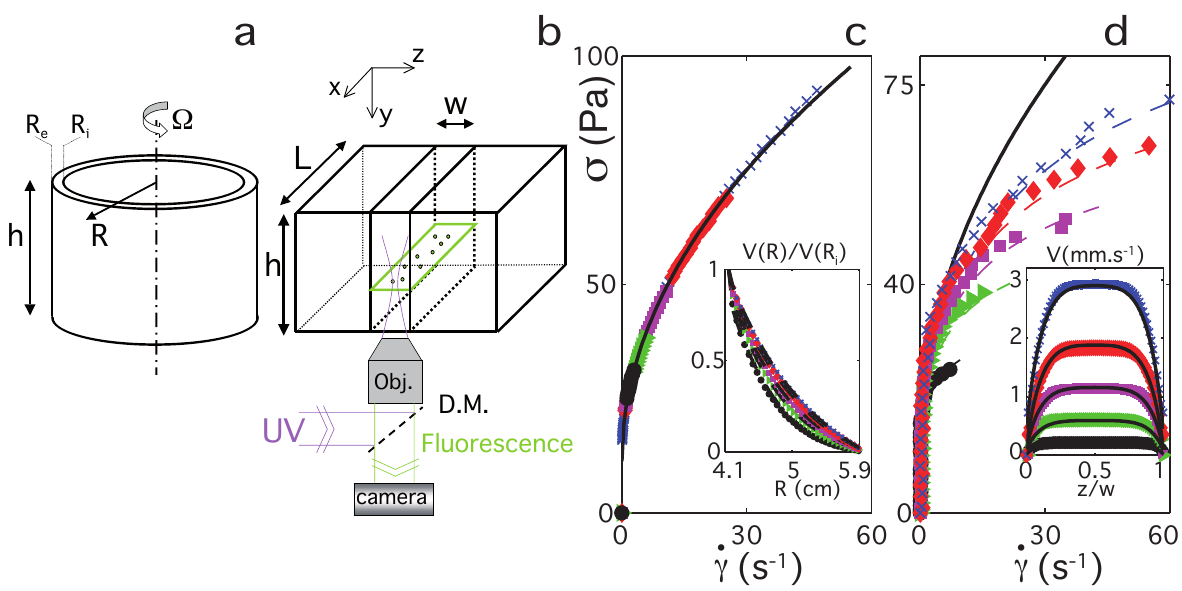}}
\end{center}
\caption{
{\bf Local flow curves measured by IRM and Microfluidic velocimetry.} \\{\it Left:} 
Schematic of the experimental set up. (A) wide gap Couette cell (inner and outer cylinder radii, $R_i=4.1$cm, and $R_e =5.9$cm respectively, height $h=11$cm). 
(B) Flow in a microchannel and particle imaging velocimetry. PIV : (DM): Dichroic mirror; UV : Ultra-Violet lamp; Obj : microscope objective. Flow direction is along $x$.
{\it Right:} Local flow curves extracted from the measurements of the velocity profiles in both geometries for an emulsion with volume fraction $\phi=0.75$ and 20\% polydispersity. The shear rates are obtained by finite differencing of the velocity profiles given in the insets. 
(C) Local flow curves, $\sigma(z)$ versus $\gammap(z)$, extracted from the velocity profiles measured within a wide gap Couette geometry (see inset). The solid line corresponds to the Herschel-Bulkley model with $\sigma_{0}=$ 11.6 Pa, $A=$ 11.2 Pa.s$^{1/2}$.
{\it Inset}: Dimensionless velocity profiles $\frac{V(R)}{V(R_i)}$ as a function of the radial coordinate $R$. From bottom to top the rotation velocity is 5, 10, 20, 50 and 100 rpm.
(D) Local flow curves, $\sigma(z)$ versus $\gammap(z)$, extracted from the velocity profiles measured in a $w=250 \mu m $ thick microchannel with rough surfaces, for various  
pressure drops as a function of the reduced coordinate $z/w$ (see inset). No overlap of the local flow curves is observed. Dashed lines are predictions for the local flow curves at the given $\Delta P$,  as obtained from the non-local rheological model of Eq. (\ref{diff}) with a cooperativity length $\xi=22.3\mu$m. The solid line corresponds to the Herschel-Bulkley model with $\sigma_{0}=$ 11.6 Pa, $A=$ 11.2 Pa.s$^{1/2}$.
{\it Inset}: Corresponding velocity profiles measured for  $\Delta P$ equal to 300, 450, 600, 750, 900 mbar. 
The length of the channel is $L=14$ cm. Solid lines are the velocity profiles predicted by the non-local rheological model of Eq. (\ref{diff}). Slippage, which is found to occur at the surfaces as detailed in Fig. 3, does not affect the resulting local flow curves as plotted in the main figure.}
\label{dispo+profil}
\end{figure*}

\begin{figure*}
\begin{center}
\scalebox{1}{\includegraphics {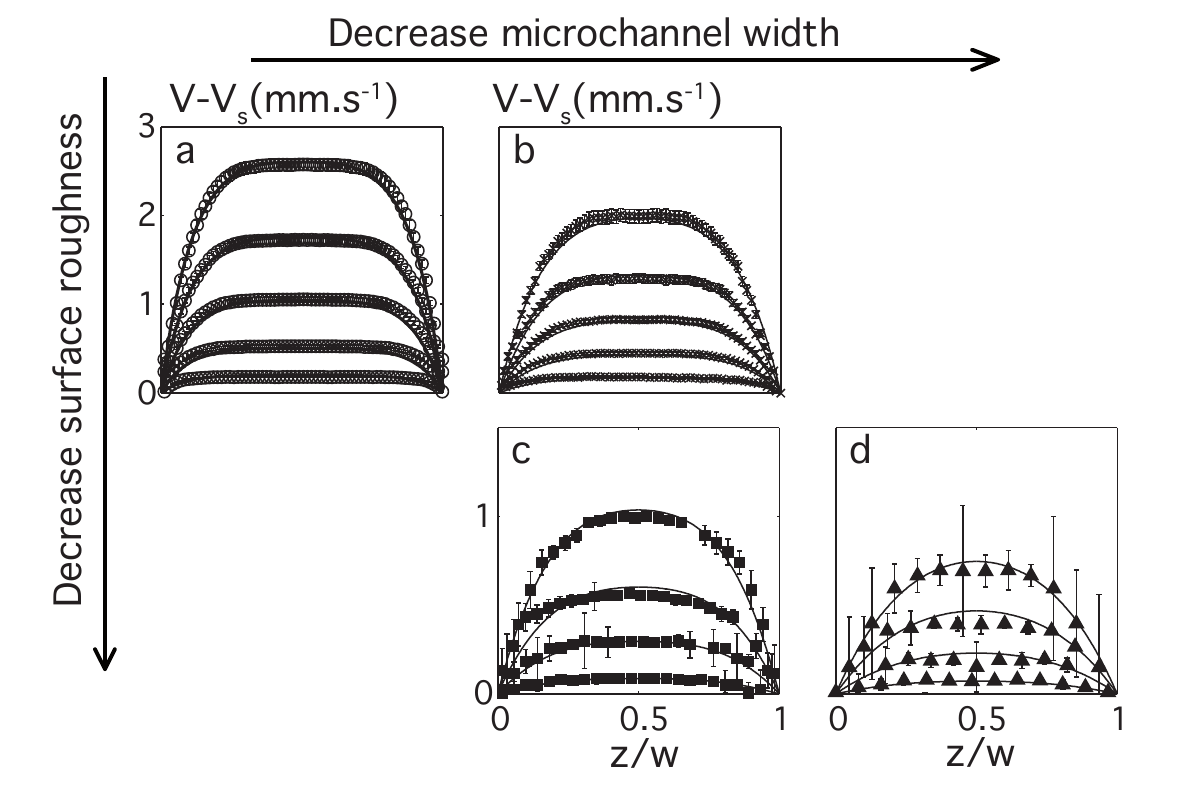}}
\end{center}
\caption{
{\bf Velocity profiles in microchannels with different gaps $w$ and confining wall roughness, corrected
for the slip velocity:} 
(a) $\circ$ rough microchannel with $w=250~\mu$m; (b) $\times$ rough microchannel with $w=125~\mu$m , (c) $\blacksquare$ smooth microchannel with $w=112~\mu$m; (d) $\blacktriangle$ smooth microchannel with $w=56~\mu$m. Solid lines are velocity profiles predicted by the non local model of Eq. (\ref{diff}) with a length $\xi=22.3\mu$m. The various experimental velocity profiles correspond to different pressure drops $\Delta P$, tuned to get the same range of  wall shear stress $\sigma_{\rm wall}$ in the various geometries. From bottom to top the values of $\sigma_{\rm wall}$ are: (a) 
27, 41, 55, 68, 82 Pa; (b) 30, 45, 60, 75, 90 Pa; 
(c) 45, 60, 75, 91 Pa; (d) 48, 65, 82, 97 Pa. 
 The volume fraction is $\phi=75\%$, with $20 \%$ polydispersity.
Errobars denote s.d. (n=450).
}
\label{courbesfittes}
\end{figure*}

\begin{figure}
\begin{center}
\scalebox{1}{\includegraphics {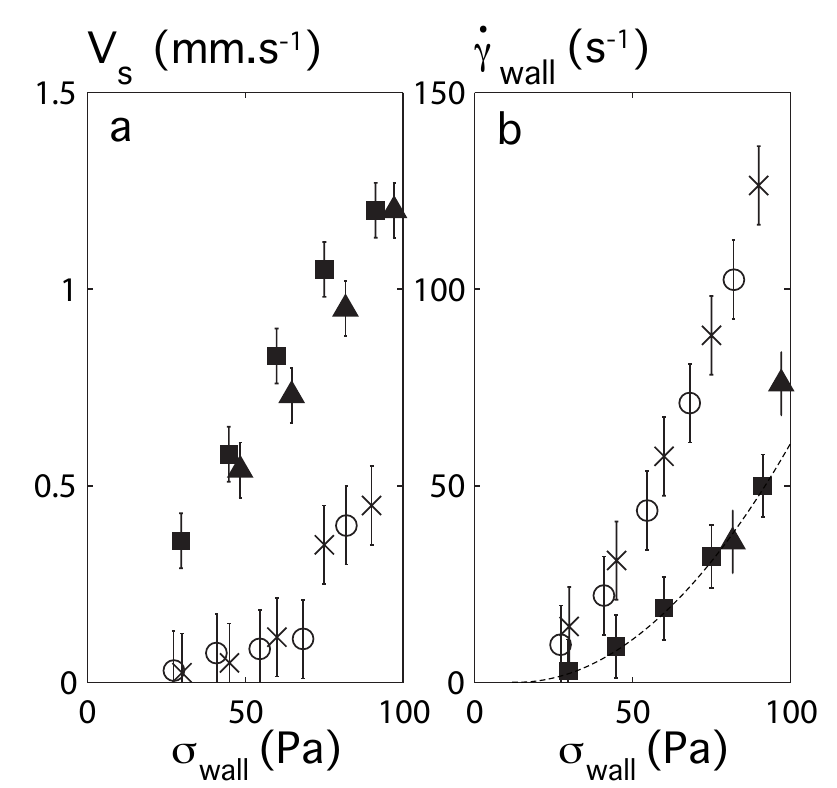}}
\end{center}
\caption{{\bf Rheology at surfaces:} (a) Slip velocity at the wall as a function of the wall shear stress for various microchannels; (b) Shear rate at the confining wall $\gammap_{\rm wall}$  as a function of the wall shear stress $\sigma_{\rm wall}$ for various microchannels. The dashed line is the bulk flow curve (Herschel-Bulkley model with $\sigma_{0}=$ 11.6 Pa, $A=$ 11.2 Pa.s$^{1/2}$).
Data gather different pressure drops $\Delta P$ and confinements $w$. Symbols are identical to those in Figure~\ref{courbesfittes}.
Open (filled) symbols correspond to rough (smooth) confining surfaces. The volume fraction is 
$\phi=75\%$, with $20 \%$ polydispersity. 
Errobars on the velocity denote s.d. (n=450).
}
%\caption{Surface rheology:  (a) Slip velocity at the wall as a function of the wall shear stress for various microchannels; (b) Shear rate at the confining wall $\gammap_{\rm wall}$  as a function of the wall shear stress $\sigma_{\rm wall}$ for various microchannels. The dashed line is the bulk flow curve (Herschel-Bulkley model with $\sigma_{0}=$ 11.65 Pa, $A=$ 11.25 SI).
%Data gather different pressure drops $\Delta P$ and confinements $w$. Symbols are identical to those in Figure~\ref{courbesfittes}.
%Open (filled) symbols correspond to rough (smooth) confining surfaces. The volume fraction is $\phi=75\%$, with $20 \%$ polydispersity. }
\label{Slip}
\end{figure}

The only remaining parameter is the bulk 'flow cooperativity length' $\xi$ in Eq. (\ref{diff}).
A crucial and surprising finding is that {\it a unique (constant) length $\xi$ accounts for all experimental data for the flow profiles and local flow curves for a given emulsion, Figs. 1.D and 2, independently of the pressure drop, confinement and surface nature} (rough or smooth). That so much data can be fitted using a single and constant length $\xi$ is quite remarkable. 
Typical comparisons are shown in Figs. \ref{dispo+profil}D and \ref{courbesfittes} for both the velocity profiles and the corresponding local flow curves  (see also Supplementary Figures SI-5, SI-6). 
{ We mention that alternative non-local rheological models, involving {\it e.g.} non-local terms in the flow curve \cite{Dhont} are unable to reproduce all measured flow profiles, Supplementary Figure SI-9.} 

\begin{figure}
\begin{center}
\scalebox{1.0}{\includegraphics {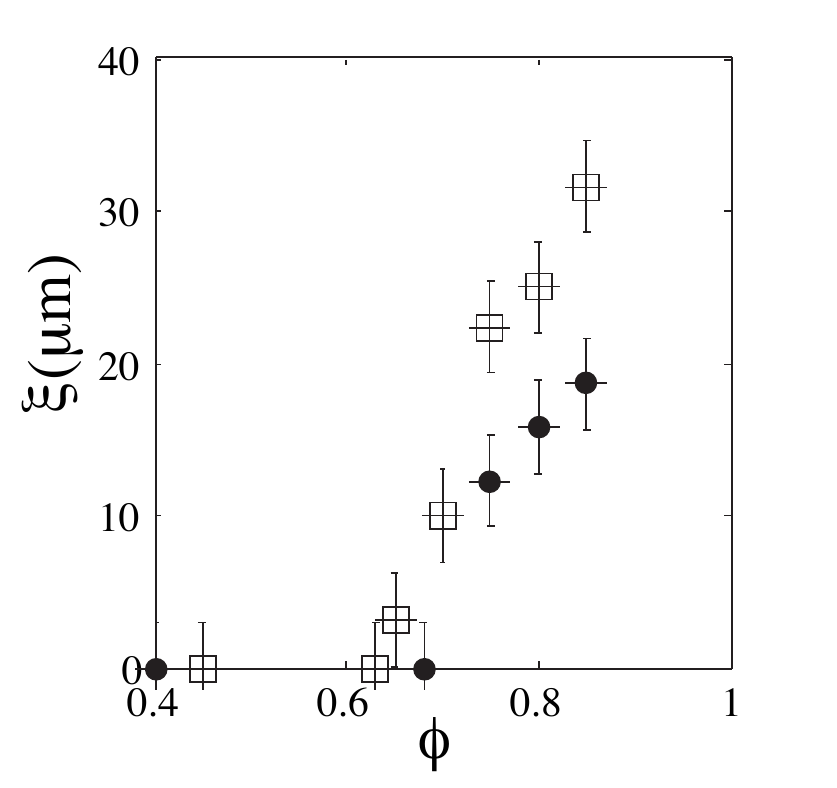}}
\end{center}
\caption{
{\bf Flow cooperativity length $\xi$ as a function of emulsion volume fraction.}
Open and filled symbols corresponds to emulsions with different polydispersities of 
$20\%$ (open) and $36\%$ (filled) respectively. The corresponding jamming volume fraction, as determined by the occurence of a non-vanishing yield stress, are $\phi_c=0.64$ (open) and $\phi_c=0.68$ (filled).
Errobars denote the dispersion of the experimental values over various experiments.
}
%\caption{Dynamical heterogeneity length $\xi$ as a function of emulsion volume fraction. 
%Open and filled symbols corresponds to emulsions with different polydispersities of 
%$20\%$ (open) and $36\%$ (filled) respectively. The corresponding jamming volume fraction, as determined by the occurence of a non-vanishing yield stress, are $\phi_c=0.64$ (open) and $\phi_c=0.68$ (filled).
%}
\label{length}
\end{figure}

Futhermore, a similar, very good agreement has been obtained for all (jammed) emulsions investigated, with various volume fractions and polydispersities (Supplementary Figures SI-5 and SI-6).
%This provides strong support for the proposed non-local rheological model. 
Fig. \ref{length} reports the variation of the measured flow cooperativity length $\xi$ as a function of the volume fraction for emulsions with two different dispersities.
Below
the jamming concentration $\phi<\phi_c$, we could not measure finite-size effects in the flow profiles within the experimental uncertainty (Supplementary Figure SI-3),  so that $\xi\approx 0$ in this case.
A strong increase of $\xi$ is observed above the jamming threshold, in line with the appearance of elastic and yield properties of the materials in this regime.
%Only above the jamming threshold $\phi>\phi_c$ is the heterogeneity length non vanishing. 
%It exhibits a steep increase with volume fraction.  

%The existence of non-local effects associated with a microscopic length are {\it a priori} not unexpected \cite{Dhont}. 
Non-local effects in the flow properties may be {\it a priori} expected at the microscopic scale  \cite{Dhont}. 
However, we measure a 'flow cooperativity length'' only in the jammed state, providing a direct and novel connection between non-local effects and the jamming transition. 
Furthermore, we emphasize that this length is associated with correlations of a dynamical quantity (the fluidity, {\it i.e.} rearrangements) and not with static structural order. 
%This further demonstrates that $\xi$ is a signature of cooperativity in the rearranging regions involved in the dynamical processes in the flow of jammed materials.
Typically  for the volume fraction range studied here, $\xi$ varies between 0 and 30 microns, 
to compare with the droplet size $d\approx 6\mu$m,
%. This value corresponds to several droplet diameters, 
giving roughly 
a cooperatively rearranging region involving up to a few hundreds of particles. These values are close to estimates for the dynamical heterogeneities measured in glassy materials \cite{Weeks,Berthier2005} or in jammed granular materials \cite{Dauchot2005,Durian2007,Bazant}. 
Furthermore, similar observations of non-locality have been reported in granular flows close to the jamming transition
\cite{Pouliquen2004,Andreotti}, suggesting further universal characteristics.

However, we emphasize that the present length characterizing flow cooperativity differs fundamentally from that which characterizes dynamical heterogeneities involved in spontaneous fluctuations. Indeed, 
while these dynamical heterogeneities 
%which grow as the transition is approached, 
%ADD
have been measured to exhibit a maximum at the jamming point \cite{Weeks,Goldman},
%ADD
the present flow cooperativity length $\xi$ is non-vanishing in the jammed region only, and 
%ADD
increases when going deeper in the jammed phase.
%$\xi$ is measured to vanish at the jamming transition ($\xi(\phi_c)=0$). 
This suggests that the flow behaviour of glassy and jammed materials involves physical mechanisms distinct from those associated with the glass transition. 
We expect that the flow cooperativity length $\xi$ measures the zone of influence of localized plastic events occuring during flow and not the size of mobile regions. 
Our results thus provides a novel perspective for developing a theoretical framework to describe the non-local flow behaviour of glassy and jammed systems, such as soft glasses and granular materials.

%{\bf \small Supplementary Information} is linked to the online version of the paper at
%www.nature.com/nature.

{\bf \small Acknowledgements}: Discussions with Jean-Louis Barrat and Bruno Andreotti are gratefully acknowledged. This project was supported by Rhodia, R\'egion Aquitaine and the ANR.
%, under program SLLOCDYN.  
L.B. acknowledges support from the von Humboldt foundation.

%{\bf \small Author Information}: Reprints and permissions information is available at npg.nature.com/reprintsandpermissions. The authors declare no competing financial interests.
%Correspondence and requests for materials should be addressed to L. B. (lyderic.bocquet@univ-lyon1.fr) and A.C. (annie.colin-exterieur@eu.rhodia.com).

\newpage

\newpage
%\begin{center}
%\scalebox{0.5}{\includegraphics {Figures/figure1}}
%\end{center}
%\newpage
%
%\begin{center}
%\scalebox{0.5}{\includegraphics {Figures/figure2}}
%\end{center}
%\newpage

%\begin{center}
%\scalebox{0.5}{\includegraphics {Figures/figure3}}
%\end{center}
%\newpage

%\begin{center}
%\scalebox{0.5}{\includegraphics {Figures/figure4}}
%\end{center}

%\newpage
%\scalebox{1.0}{\includegraphics {Figures/figure1}}
%\newpage
%\scalebox{1}{\includegraphics {Figures/figure2}}
%\newpage
%\scalebox{1}{\includegraphics {Figures/figure3}}
%\newpage
%\scalebox{1.0}{\includegraphics {Figures/figure4}}

 \end{document}